\documentclass[prd,preprint,aps,showpacs,tightenlines,eqsecnum]{revtex4}

\usepackage{amsmath}
\usepackage{amssymb}

\newcommand{\beq}{\begin{equation}}
\newcommand{\eeq}{\end{equation}}
\newcommand{\beqa}{\begin{eqnarray}}
\newcommand{\eeqa}{\end{eqnarray}}

\begin{document}

\title{On the FP-ghost propagators for Yang-Mills theories and perturbative 
quantum gravity in the covariant gauge
in de~Sitter spacetime}

\author{ Mir Faizal$^1$ and Atsushi Higuchi$^2$}

\affiliation{Department of Mathematics, University of York,
Heslington, York YO10 5DD, UK\\ email: $^1$fm521@york.ac.uk,
$^2$ah28@york.ac.uk}

\date{September 8, 2008}

\pacs{04.62.+v, 04.60.-m}

\begin{abstract} 
The propagators of the Faddeev-Popov (FP)
ghosts for Yang-Mills theories and perturbative quantum gravity in
the covariant gauge are infrared (IR) divergent in 
de~Sitter spacetime.  We point out, however, that the modes responsible for 
these divergences will not contribute to loop diagrams in
computations of time-ordered products in either Yang-Mills theories
or perturbative quantum gravity.  Therefore we propose that the IR divergent FP-ghost propagator should be regularized by a
small mass term that is sent to zero in the end of any perturbative calculations.    
This proposal is equivalent to using the effective FP-ghost
propagators, which we present in an explicit form, obtained by removing the modes responsible for the IR divergences.  
We also make some comments on the corresponding propagators in anti-de~Sitter spacetime.
\end{abstract}

\maketitle

\section{Introduction}
Quantum field theory in de~Sitter spacetime~\cite{1} has been actively
studied recently due to its relevance to 
inflationary cosmologies~\cite{Guth,Linde,Steinhardt}. 
Furthermore, the current observations indicate that our Universe is 
expanding in an accelerated rate and may approach de~Sitter spacetime
asymptotically~\cite{Riess,Perlmutter}.
In order to study higher-order quantum effects for Yang-Mills theories or
perturbative gravity, one needs to introduce Faddeev-Popov
(FP) ghosts except in unwieldy gauges such as the axial gauge (for the
case of Yang-Mills theories).  In this paper we study the Feynman
propagators for the FP ghosts in these theories in de~Sitter spacetime.

Let us emphasize that there is nothing wrong with 
covariantly-quantized linearized
gravity in de~Sitter spacetime~\cite{AllenTuryn,HKcovariant} in spite of the recent claim to the 
contrary by Woodard~\cite{Woodard}, who maintains that even the
retarded Green's function fails to work in this theory.  This controversy is partly due to
the gauge chosen by the well-known work of Antoniadis and
Mottola~\cite{AntoniadisMottola} that introduces spurious infrared (IR)
divergences~\cite{Allen86b}.  Another source of confusion is
that the spacelike nature of the past infinity of de~Sitter spacetime
necessitates the inclusion of the initial data on the past infinity in
calculating the field using the retarded Green's function.
In fact the covariant retarded Green's 
function obtained in Ref.~\cite{HKcovariant} reproduces the linearized gravitational field from
static point masses if the initial data on the
spacelike past infinity is taken into account~\cite{HiguchiLee}.

Unlike the propagators for the gauge fields and linearized gravitational fields, 
the FP-ghost propagators for these theories are indeed IR divergent. 
However, the interaction between the Yang-Mills/gravitational field and the FP ghosts is such that,
if we regularize the IR divergences by introducing a small mass term, the modes
responsible for the IR divergences will not contribute in the
computation of time-ordered products of physical fields.
For this reason, we propose that one should regularize the IR divergences of 
the FP-ghost propagators in these theories
and then take the limit where the regularization is removed.
This proposal is equivalent to using the effective Feynman propagators obtained by subtracting 
the modes responsible for IR divergence in perturbative calculations.  
In this paper we present these effective FP-ghost propagators.

In the rest of this paper we treat the Yang-Mills
case in Sec.~\ref{sec2} and the perturbative-gravity case in
Sec.~\ref{sec3}, and we make
some comments on the corresponding FP-ghost
propagators in anti-de~Sitter spacetime in Sec.~\ref{sec4}.
Our metric signature is $-+++$.

\section{Yang-Mills Theories} \label{sec2}
The metric for $4$-dimensional 
de~Sitter spacetime is given by
\begin{equation}
ds^2 = -dt^2 + H^{-2}\cosh^2Ht\left( d\chi^2 +
\sin^2\chi\,d\Omega_2^2\right),
\end{equation}
where $d\Omega_2^2$ is the metric on the unit $2$-sphere and where 
$0\leq \chi < \pi/2$.  We let $H=1$ from now on for simplicity.
We consider the Yang-Mills theory in this spacetime with 
the gauge group with structure constant $f^a_{bc}$ and the gauge field
$A_\mu^a$. The gauge-fixing term in the Lagrangian density is
\begin{equation}
{\cal L}_{\rm gf} = \sqrt{-g}\left[ 
-B^a\nabla_\mu A^{a\mu} + \frac{\alpha}{2}B^a B^a\right],
\end{equation}
whereas the Faddeev-Popov term is
\begin{equation}
{\cal L}_{\rm PF} = i\sqrt{-g}\nabla^\mu \overline{c}^a D_\mu c^a
\end{equation}
with $D_\mu c^a = \nabla_\mu c^a + f^a_{bd}A_\mu^b c^d$. 
Here, $g$ denotes the determinant of the metric $g_{\mu\nu}$ on de~Sitter spacetime.  
The total Lagrangian density is
${\cal L} = {\cal L}_{\rm cl} + {\cal L}_{\rm gf} + {\cal L}_{\rm FP}$,
where ${\cal L}_{\rm cl}$ is the classical Lagrangian density for the
Yang-Mills field.

The non-interacting part of the FP-ghost Lagrangian density is
$i\nabla^\mu\overline{c}^a\nabla_\mu c^a$.  Thus, the FP ghosts are
minimally-coupled massless scalar fields, which are known to suffer from IR
divergences~\cite{FordParker,Ratra,Allen}.  If FP-ghosts were physical fields, one would need to break de~Sitter invariance of the vacuum for these fields~\cite{Allen,AllenFolacci}.
However, as we shall see, this problem can be circumvented
because they are unphysical fields appearing only in internal loops in Feynman diagrams and 
couple to the gauge field only through a derivative coupling.

Following Allen and Jacobson~\cite{AllenJacobson} we define
$\mu(x,x')$ to be the geodesic distance between spacelike-separated 
points $x$ and $x'$ in de~Sitter spacetime
define the variable $z=\cos^2(\mu/2)$. In view of
the IR divergences mentioned above, we first consider the
propagator defined by 
\begin{equation}
T\langle c^a(x)\overline{c}^b(x')\rangle
= i\delta^{ab}D_{m^2}(x,x')
\end{equation}
for the FP ghosts with small mass $m$ to regularize the IR divergences. 
As is well known, in the so-called the Euclidean vacuum~\cite{GibbonsHawking}, 
one has~\cite{BunchDavies,AllenJacobson}
\begin{eqnarray}
D_{m^2}(x,x') & = & \frac{1}{16\pi^2}\Gamma(a_{+})
\Gamma(a_{-}) F(a_{+}, a_{-}; 2; z) \nonumber \\
& = & \frac{1}{16\pi^2}\sum_{n=0}^\infty \frac{\Gamma(a_{+}+n)\Gamma(a_{-}+n)}{(n+1)!n!}z^n, 
\label{hgeo}
\end{eqnarray}
where 
$a_{\pm} = \frac{3}{2} \pm \left( \frac{9}{4} - m^2 \right)^{\frac{1}{2}}$
and where $F(\alpha,\beta;\gamma;z)$ is Gauss' hypergeometric function.  The function $D_{m^2}(x,x')$ is defined for non-spacelike separated points $x$ and $x'$ by a suitable analytic continuation.

In the limit $m\to 0$ we have $a_{-}\to 0$.  Hence 
the first term in the series expansion (\ref{hgeo}) gives a $z$-independent IR-divergent contribution.
Now, the interaction term involving the FP ghosts is 
$if^a_{bd}\nabla^\mu \overline{c}^aA^b_\mu c^d$.  Since the FP ghosts
appear only in internal loops and couple to the gauge field through a
derivative coupling, the first term in the hypergeometric series 
(\ref{hgeo}) does not contribute to the calculation of
$n$-point functions of the gauge fields.  Therefore, we propose that one should use the
effective FP-ghost propagator obtained by subtracting this
contribution.  Thus we subtract $1$ from $F(a_+,a_-;2;z)$ in
Eq.~(\ref{hgeo}), take the limit $m\to 0$ and add any constant term 
to obtain the effective FP-ghost propagator as
\begin{equation}
D_0^{\rm eff}(x,x') = 
\frac{1}{16\pi^2}\left[\frac{1}{1-z} - 2\log(1-z) + C\right],
\end{equation}
where $C$ is an arbitrary constant.
This effective propagator with $C = -14/3$
was used in calculating the covariant graviton
propagators~\cite{AllenTuryn,HKcovariant}.

%

\section{Perturbative gravity} \label{sec3}

Since the FP ghosts for perturbative gravity are vector fields, we need
to review the formalism of Allen and Jacobson for the vector propagators
in maximally-symmetric spaces or spacetimes~\cite{AllenJacobson}.   Let $x$ and
$x'$ be two spacelike separated points and let $\mu(x,x')$ be the
geodesic distance between them as before.  One defines the unit tangent
vectors $n_\alpha$ at $x$ and $n_{\alpha'}$ at $x'$ along the geodesic between
these two points by $n_\alpha = \nabla_\alpha \mu(x,x')$, where the
differentiation is with respect to $x$, and $n_{\mu'} =
\nabla_{\alpha'}\mu(x,x')$, where the differentiation is with respect to
$x'$.  In addition one defines the parallel propagator
$g_{\alpha\alpha'}(x,x')$ such that if $V^\alpha$ is a vector at $x$,
then $V^{\alpha'} = V^\alpha {g_\alpha}^{\alpha'}$ is the vector at $x'$
obtained by parallelly transporting $V^\alpha$ along the geodesic. 
Then, $V^\alpha = {g^\alpha}_{\alpha'}V^{\alpha'}$.  One also writes the
metric tensors at $x$ and $x'$ as $g_{\mu\nu}$ and $g_{\mu'\nu'}$,
respectively.  Any covariant bi-vectors in a
maximally-symmetric space(time) 
such as de~Sitter spacetime can be expressed as 
$\alpha(z)g_{\mu\mu'} + \beta(z)n_\mu n_{\mu'}$. 

In perturbative gravity one writes the full metric as 
$g^{\rm (f)}_{\mu\nu} = g_{\mu\nu} + h_{\mu\nu}$, where $g_{\mu\nu}$ is
the metric of the background spacetime and where $h_{\mu\nu}$ is
regarded as small.  The covariant gauge-fixing term is
\begin{equation}
{\cal L}_{\rm gf} = \sqrt{-g}\left[ 
- B^\mu(\nabla^\nu h_{\mu\nu} - k \nabla_\mu
{h^{\nu}}_\nu) + \frac{\alpha}{2}B^\mu B_\mu\right],
\end{equation}
where the covariant derivative is the one compatible with the background
de~Sitter spacetime and where $k$ and $\alpha$ are gauge parameters.  
The indices are lowered and raised by $g_{\mu\nu}$. 
The infinitesimal gauge transformation is given by
\begin{equation}
\delta_\Lambda h_{\mu\nu}  
 =  \nabla_\mu \Lambda_\nu + \nabla_\nu \Lambda_\mu
+ \pounds_{\Lambda}h_{\mu\nu},
\end{equation}
where
\begin{equation}
(\pounds_\Lambda h)_{\mu\nu}
= \Lambda^\alpha \nabla_\alpha h_{\mu\nu}
+ h_{\alpha\nu}\nabla_\mu \Lambda^\alpha + h_{\mu\alpha}\nabla_\nu
\Lambda^\alpha 
\end{equation}
is the Lie derivative of $h_{\mu\nu}$ with respect to the vector field
$\Lambda^\mu$.  Hence, the FP-ghost term in the Lagrangian density up to a total derivative is
\begin{eqnarray}
{\cal L}_{\rm FP} & = & -i\sqrt{-g}\,\overline{c}^\mu \delta_c
(\nabla^\nu h_{\mu\nu} - k\nabla_\mu {h^{\nu}}_\nu)\nonumber \\
& = & i \sqrt{-g}
\nabla^\mu \overline{c}^\nu \left[\nabla_\mu
c_\nu +
\nabla_\nu c_\mu - 2k g_{\mu\nu}
\nabla_\beta c^\beta \right.\nonumber \\
&& \left. +  \pounds_c h_{\mu\nu} - k g_{\mu\nu}g^{\alpha\beta}
\pounds_c h_{\alpha\beta}\right]. \label{FPlag}
\end{eqnarray}
The total Lagrangian density is ${\cal L}_{\rm GR} + {\cal L}_{\rm gf} +
{\cal L}_{\rm FP}$, where ${\cal L}_{\rm GR}$ is the Enstein-Hilbert
action with a positive cosmological constant.  
We are interested only in ${\cal L}_{\rm FP}$ in
this paper.

The free field equation for the FP ghost $c_\mu$ can be written as
\begin{equation}
\nabla^\mu (\nabla_\mu c_\nu + \nabla_\nu c_\mu - 2k
g_{\mu\nu}\nabla^\alpha c_\alpha) = 0,
\end{equation}
and the anti-ghost $\overline{c}_\mu$ satisfies the same equation.  On
$S^4$, where $R_{\mu\nu} = 3g_{\mu\nu}$, this equation can
be written as
\begin{equation}
\nabla^\mu(\nabla_\mu c_\nu - \nabla_\nu c_\mu)
- 2\beta^{-1}\nabla_\nu (\nabla_\alpha c^\alpha) + 6c_\nu = 0,
\end{equation}
where we have defined
$k = 1 + 1/\beta$.
Let us write the ghost propagator as
\begin{equation}
T\langle c_\mu(x)\overline{c}_{\nu'}(x')\rangle = iG_{\mu\nu'}(x,x').
\end{equation}
The function $G_{\mu\nu'}(x,x')$ is the unique bi-vector function on $S^4$
satisfying
\begin{equation}
{L_\mu}^\nu G_{\nu\nu'}(x,x') = g_{\mu\nu'}\delta^4(x,x'), \label{Green}
\end{equation}
where
\begin{equation}
{L_\mu}^\nu = - \delta_\mu^\nu \nabla_\alpha\nabla^\alpha +
\nabla^\nu \nabla_\mu + 2\beta^{-1} \nabla_\mu \nabla^\nu -
6\delta_\mu^\nu.
\end{equation}

We use the fact the Feynman propagators in the Euclidean vacuum 
in de~Sitter spacetime can be 
obtained from the corresponding Green's functions on the 4-sphere~\cite{AllenTuryn}, 
which is the Euclidean section of de~Sitter spacetime, obtained by 
the transformation $\tau = \pi/2 -it$.
Any smooth vector field on $S^4$ can be expressed as a linear combination
of the divergence-free vectors $V_\mu^{(n,\sigma)}(x)$, and
the gradient, $\nabla_\mu \phi^{(n,\sigma)}(x)$, $n=1,2,\ldots$, where
\begin{eqnarray}
-\nabla^\mu\nabla_\mu\phi^{(n,\sigma)} 
& = & n(n+3)\phi^{(n,\sigma)},\,\,\,n=0,1,2,\ldots,\\
- \nabla^\nu (\nabla_\nu V_\mu^{(n,\sigma)} - \nabla_\mu
V_\nu^{(n,\sigma)}) & = &  (n+1)(n+2)V_\mu^{(n,\sigma)},
\end{eqnarray}
with
\begin{eqnarray}
\int_{S^4}dx^4 \overline{\phi^{(n,\sigma)}}\phi^{(n',\sigma')} & = &
\delta^{nn'}\delta^{\sigma\sigma'},\\
\int_{S^4} d^4x \overline{V_\mu^{(n,\sigma)}}V^{(n',\sigma')\mu} & = &
\delta^{nn'}\delta^{\sigma\sigma'},
\end{eqnarray}
where $\sigma$ denotes the labels other than $n$.
The vector delta-function on the right-hand side of Eq.~(\ref{Green})
can be expressed as
\begin{equation}
g_{\mu\nu'}\delta^4(x,x') = \delta^{(V)}_{\mu\nu'}(x,x') +
\delta^{(S)}_{\mu\nu'}(x,x'),
\end{equation}
where
\begin{eqnarray}
\delta^{(V)}_{\mu\nu'}(x,x') & = & \sum_{n=1}^\infty \sum_\sigma
V_{\mu}^{(n,\sigma)}(x) \overline{V_{\nu'}^{(n,\sigma)}(x')},\\
\delta^{(S)}_{\mu\nu'}(x,x') & = & 
\sum_{n=1}^\infty \sum_\sigma \frac{1}{n(n+3)}\nabla_\mu \phi^{(n,\sigma)}(x)
\nabla_{\nu'} \overline{\phi^{(n,\sigma)}(x')}. 
\end{eqnarray}
We look for the Green's function in the form
\begin{equation}
G_{\mu\nu'}(x,x') = G^{(V)}_{\mu\nu'}(x,x') + G^{(S)}_{\mu\nu'}(x,x'),
\end{equation}
where 
\begin{eqnarray}
G^{(V)}_{\mu\nu'}(x,x') & = & 
\sum_{n=1}^\infty \sum_\sigma
c^{(V)}_n\,V_{\mu}^{(n,\sigma)}(x) \overline{V_{\nu'}^{(n,\sigma)}(x')},\\
G^{(S)}_{\mu\nu'}(x,x') & = & 
\sum_{n=0}^\infty \sum_\sigma c^{(S)}_n\,\nabla_\mu \phi^{(n,\sigma)}(x)
\nabla_{\nu'} \overline{\phi^{(n,\sigma)}(x')}.
\end{eqnarray}
Eq.~(\ref{Green}) is solved by
\begin{eqnarray}
c^{(V)}_n & = & \frac{1}{(n+1)(n+2)-6},\label{cVn}\\
c^{(S)}_n & = & -\frac{1}{6}\left[ \frac{1}{n(n+3)} -
\frac{1}{n(n+3)+3\beta}\right]. \label{cSn}
\end{eqnarray}
We find from the expression of $c^{(V)}_n$ that the Green's
function $G^{(V)}_{\nu\nu'}(x,x')$ is IR divergent because of the
contribution from the $n=1$ modes, which are
Killing vectors.  Since the coupling term of the FP ghosts to the
metric perturbation $h_{\mu\nu}$ in Eq.~(\ref{FPlag}) is proportional to
$\nabla_\mu \overline{c}_\nu + \nabla_\nu\overline{c}_\mu$,
the $n=1$ modes do not contribute to
loop diagrams. Hence, as in the Yang-Mills case,  
we propose that one should use an effective propagator obtained by
subtracting this IR-divergent contribution.

The propagator for the divergence-free vector field of
arbitrary mass has been given by Allen and
Jacobson~\cite{AllenJacobson}.  Let
\begin{equation}
\gamma(z) = - \frac{3\Gamma(b_+)\Gamma(b_-)}{64\pi^2
m^2}F(b_+,b_-;3;z), \label{hyper}
\end{equation}
where $b_\pm = \frac{5}{2} \pm \frac{1}{2}(1-4m^2)^{1/2}$.
(Here, $m=0$ corresponds to the gauge theory.) Then
\begin{eqnarray}
G_{\nu\nu'}^{(V)}(x,x') & = & \alpha^{(V)}(z)g_{\nu\nu'} +
\beta^{(V)}(z)n_{\nu}n_{\nu'} \nonumber \\
&& - \frac{1}{m^2}\nabla_{\nu}\nabla_{\nu'}D_0^{\rm eff}(x,x'), \label{well}
\end{eqnarray}
where
\begin{eqnarray}
 \alpha^{(V)} (z) & = & \left[-\frac{2}{3}z(1 - z )\frac{d}{dz} + 
2z-1 \right] \gamma(z),\label{alpha}\\
\beta^{(V)}(z) & = & \alpha^{(V)}(z) - \gamma(z). \label{beta}
\end{eqnarray}

The vector part of the FP ghosts
satisfies the massive vector equation with $m^2=-6$.
Since one has~\cite{AllenTuryn}
\begin{eqnarray}
&& \sum_{\sigma}V^{(1,\sigma)}(x)\overline{V^{(1,\sigma)}(x')}\nonumber
\\
&& =
\frac{15}{16\pi^2} \left[(2z-1)g_{\nu\nu'} + 2(z-1)n_\nu
n_{\nu'}\right],
\end{eqnarray}
the infinite contribution due to the $n=1$ modes, which do not
contribute to ghost-loop diagrams, comes from the $z$-independent part
of $\gamma(z)$ in Eq.~(\ref{hyper}). 
Hence we may let
\begin{eqnarray}
\gamma^{\rm eff}(z) & = & -\lim_{m^2\to -6}
\frac{3\Gamma(b_+)\Gamma(b_-)}{64\pi^2
m^2}\left[F(b_+,b_-;3;z)-1\right] + C \nonumber \\
& = & \frac{1}{64\pi^2}\left[ \frac{1}{(1-z)^2} + \frac{6}{1-z} -
12\log(1-z)+4\right],
\label{theresult}
\end{eqnarray}
where we have chosen $C=11/64\pi^2$.
The effective functions $\alpha^{(V)}(z)$ and $\beta^{(V)}(z)$ in Eq.~(\ref{well})
are given by substituting Eq.~(\ref{theresult}) in
Eqs.~(\ref{alpha}) and (\ref{beta}).

The scalar contribution found from Eq.~(\ref{cSn}) is
\begin{equation}
 G^{(S)}_{\nu\nu'}(x,x') = -\frac{1}{6}\nabla_\nu \nabla_{\nu'}[D_0^{\rm
eff}(x,x') - D_{3\beta}(x,x')]. \label{scalarpart}
\end{equation}
One can see from Eq.~(\ref{cSn}) that 
the Green's function $D_{3\beta}(x,x')$ is IR divergent for
$3\beta = -N(N+3)$, $N=1,2,3,\ldots$.
The covariant graviton propagator is IR divergent for the same values of $\beta$~\cite{Allen86b}.  
The graviton propagator used by Antoniadis
and Mottola~\cite{AntoniadisMottola} corresponds to the $N=1$ case and
is IR divergent as a result.
The propagator $G^{(S)}_{\nu\nu'}$ takes a simple form for $\beta =
2/3$ because $D_{2}(x,x') = (16\pi^2)^{-1}(1-z)^{-1}$, which is the
propagator for the conformally-coupled massless scalar field.
We can find $G^{(S)}_{\nu\nu'}$
by using the formula~\cite{AllenTuryn}
\begin{equation}
\nabla_\nu\nabla_{\nu'}f(z) = \tfrac{1}{2}f'(z)g_{\nu\nu'}+
(1-z)\left[zf'(z)\right]'n_\nu n_{\nu'}.  \label{useful}
\end{equation}
Adding $G^{(V){\rm eff}}_{\nu\nu'}(x,x')$ and
$G^{(S)}_{\nu\nu'}(x,x')$, we have
\begin{equation}
G^{\rm eff}_{\nu\nu'}(x,x')
= \alpha^{\rm eff}(z)g_{\nu\nu'} + \beta^{\rm eff}(z)n_{\nu}n_{\nu'},
\end{equation}
where, for $\beta=2/3$,
\begin{eqnarray}
\alpha^{\rm eff}(z) & = & \frac{1}{16\pi^2}
\left[ \frac{1}{3(1-z)} - 3 -3(2z-1)\log(1-z)\right],\\
\beta^{\rm eff}(z) & = & 
\frac{1}{16\pi^2}\left[-\frac{4}{3(1-z)} 
- 4 +6(1-z)\log(1-z)\right].
\end{eqnarray}

\section{Comments on the anti-de~Sitter case} \label{sec4}

Let us make some comments on the FP-ghost propagators 
in anti-de~Sitter spacetime
since this spacetime has attracted much attention recently because of
the AdS/CFT correspondence~\cite{Maldacena,Polyakov,Witten}.
The propagators for the FP ghosts in anti-de~Sitter spacetime in $n$
dimensions can readily be obtained using the work of Allen and
Jacobson~\cite{AllenJacobson} since there is no IR problem unlike
in the de~Sitter case. 
In the $n$-dimensional anti-de~Sitter spacetime 
there is some freedom in the boundary condition at spatial infinity if 
the mass $m$ of the minimally-coupled scalar field satisfies
$-(n-1)^2/4 < m^2 < - (n-1)^2/4 + 1$~\cite{Avis,Freedman}.
However, there is only one possible boundary condition for
the FP ghosts for Yang-Mills theories since they are minimally-coupled
massless scalar fields with $m^2=0$.  
One can readily show that this is also the case
for the divergence-free 
part of the FP ghosts for perturbative gravity. (The condition for more
than one possible boundary condition for the divergence-free vector modes
is $-(n-3)^2/4 < m^2 < -(n-3)^2/4+1$. One has $m^2 =n$ for the FP ghosts.) The
mass of the scalar part depends on the gauge parameter $\beta$.  This
part of the propagator is obtained by replacing $D_0^{\rm
eff}(x,x')-D_{3\beta}(x,x')$ in Eq.~(\ref{scalarpart}) by 
$\Delta_0(x,x') - \Delta_{-3\beta}(x,x')$, where $\Delta_{m^2}$ is the
propagator for the minimally-coupled 
scalar field of mass $m$ in anti-de~Sitter spacetime. There is some
freedom in the choice of boundary condition if $(n-1)^2/4 - 1 < 3\beta
< (n-1)^2/4$.  The propagator $\Delta_{-3\beta}(x,x')$
given by Allen and Jacobson~\cite{AllenJacobson} satisfies the boundary
condition such that it falls off as rapidly as possible for $z\to\infty$.
The case $3\beta=n(n-2)/4$ corresponds to the conformally-coupled
massless scalar field.

\end{document}